\documentclass[12 pt]{article}
\usepackage[utf8]{inputenc}
\usepackage{geometry}
\title{WRI-176 PD-1}
\author{Areeq Hasan}
\date{April 16, 2021}

\usepackage{fancyhdr}
\usepackage{extramarks}
\usepackage{amsmath}
\usepackage{amssymb}
\usepackage{amsthm}
\usepackage{amsfonts}
\usepackage[shortlabels]{enumitem}
\usepackage{booktabs}
\usepackage{graphicx}
\usepackage{hyperref}
\usepackage{biblatex}

\providetoggle{blx@lang@captions@<language>}
\newcommand{\course}[0]{\noindent Princeton University\\}
\newcommand{\prof}[0]{\noindent Areeq I. Hasan\\}
\newcommand{\wordcount}[0]{\noindent May 26, 2021}
\renewcommand{\title}[0]{\noindent IGO-QNN: Quantum Neural Network Architecture for Inductive Grover Oracularization}
\begin{document}

\setcounter{page}{0}

\thispagestyle{fancy}

    \prof
    \course
    \wordcount

	\centering
	\vspace{1.5cm}

	{\scshape\Large \title\par}
	\vspace{1.5cm}
	\raggedright

\setlength{\parindent}{5ex}

\section*{Abstract}
We propose a novel paradigm of integration of Grover's algorithm in a machine learning framework: the inductive Grover oracular quantum neural network (IGO-QNN). The model defines a variational quantum circuit with hidden layers of parameterized quantum neurons densely connected via entangle synapses to encode a dynamic Grover's search oracle that can be trained from a set of database-hit training examples. This widens the range of problem applications of Grover's unstructured search algorithm to include the vast majority of problems lacking analytic descriptions of solution verifiers, allowing for quadratic speed-up in unstructured search for the set of search problems with relationships between input and output spaces that are tractably underivable deductively. This generalization of Grover's oracularization may prove particularly effective in deep reinforcement learning, computer vision, and, more generally, as a feature vector classifier at the top of an existing model.

\pagebreak

\tableofcontents

\pagebreak

\section{Introduction}
Quantum computing has recently been demonstrated to host space-wise and time-wise computational complexity advantages that have the potential to significantly improve machine learning performance,\cite{cai_entanglement-based_2015} particularly in the context of neural networks.\cite{liu_single-hidden-layer_2013}\cite{dunjko_machine_2018}\cite{clark_basis_2014}\cite{beer_training_2020}\cite{skolik_layerwise_2021} One algorithm that utilizes these paradigmatic advantages to achieve quadratic speedup in unstructured search is Grover's search. Grover's unstructured search algorithm has been explored as an optimizer for finding loss-minimizing model parameters in k-medians and k-nearest neighbors algorithms as well as single-hidden layer quantum neural networks,\cite{liu_single-hidden-layer_2013}\cite{du_grover-search_2021} but it has not been used for multi-hidden layer quantum neural networks (QNNs). Furthermore, existing architectures use single qubit gates exclusively for quantum neurons which do not take advantage of quantum entanglement for solving problem archetypes that are more conducive to the medium. Most significantly, an explicit oracle encoding a solution verification algorithm is required for Grover's search narrowing its domain to search problems with known verifiers. We propose a novel paradigm of integration of Grover's algorithm in a machine learning framework: the inductive Grover oracular quantum neural network (IGO-QNN). The model defines a variational quantum circuit with hidden layers of parameterized quantum neurons densely connected via entangle synapses to encode a dynamic Grover's search oracle that can be trained from a set of database-hit training examples. This widens the range of problem applications of Grover's unstructured search algorithm to include the vast majority of problems lacking analytic descriptions of solution verifiers, allowing for quadratic speed-up in unstructured search for the set of search problems with dependency relationships that are tractably underivable via deductive means. This generalization of Grover's oracularization may prove particularly effective in deep reinforcement learning, computer vision, and, more generally, as a feature vector classifier at the top of an existing model. The rest of this paper is organized as follows. First, the paper provides background on classical neural networks, advantages of the quantum computational paradigm, and Grover’s unstructured search algorithm. The paper then addresses and analyzes existing works related to integrating Grover’s search in a machine learning context. The background and related works are syncretized in the motivation section before the paper discusses the model architecture, network propagation behavior, and model training, in-depth. Finally, the paper addresses applications of the proposed model.

\section{Background}
\subsection{Classical Neural Networks}
In the analytical computational paradigm, an input and a set of rules (that maps input to output) are taken in and used to deductively generate output corresponding to the given input. In the machine learning paradigm, input and output are taken in and used to inductively generate a set of rules mapping from input to output. 

A classical neural network is a machine learning model that applies this inductive computational paradigm to learn complex relationships between input and output layers by iteratively tuning synaptic weights and neural biases. The model resembles a directed graph of neural nodes and synaptic edges. The input vector is densely connected to each neuron in the adjacent hidden layer $\ell_1$, each neuron in layer $\ell_i$ is densely connected to each neuron in $\ell_{i+1}$, and the neurons in layer $\ell_\ell$ is densely connected to the output vector. Dense connections are encoded via graph edges that represent a weighted sum of edges in forward propagation through the network. Each neuron in a hidden layer is transformed by a non-linear activation function such that complex feature crosses can be learned by the network. Thus, a single pass through the model with a given input vector returns an output vector that is determined by the configuration of weights and biases. These model parameters are tuned to a given mapping between input-output vectors via the process of \textit{model training} where an optimizer algorithm iteratively forward propagates a given input from a training example (for which the output is known), computes the loss of the predicted output from the expected output, and backpropagates through the network to tweak synaptic weights and neural biases to minimize the loss of the model. 

Deep classical neural networks have proven particularly effective in solving problems with complex relationships between input and output vectors that are near impossible to effectively encode using an analytically-derived algorithm. Sufficiently complex tasks, particularly in artificial intelligence subdomains of reinforcement learning and natural language processing, require classical neural networks with billions of parameters that can be intractable to train efficiently even in parallel with optimizations such as network weight-pruning. 
However, recent advances in quantum computational complexity theory have indicated significant potential advantages of robust quantum machine learning (QML) models, specifically quantum neural networks, over classical models.

\subsection{The Quantum Advantage}

Quantum computing revolves around the qubit, the unit of quantum information. Quantum properties extend the degrees of freedom of the qubit such that its state can be represented as a complex vector in its Hilbert space where the computational basis states are the parallel and antiparallel polar axes of the corresponding Bloch sphere, and quantum gates perform unitary transformations that rotate the wave function statevector to any point on the sphere. If the qubit statevector points in between the computational basis states, its wave function is in superposition and this allows for exponential space-wise efficiency gains in the quantum representation of information | Hadamard transforms allows for the encoding of $2^n$ booleans where $n$ is the number of qubits; that is, the same system would require exponentially more space to be represented classically. To retrieve information from the qubit, it must be measured to a classical register which collapses the statevector to point along one of the computational bases. While these extra degrees of freedom in the representation of the information are lost in measurement, delaying measurement to the end of quantum algorithm clever use of quantum properties can allow for significant time-wise efficiency gains. Instances of this delay in measurement include Shor’s algorithm, which implements quantum fourier transform and period finding to tractably factor large numbers, and Grover’s algorithm, which allows for quadratic speed-up in unstructured search by employing quantum parallelism and oracularization. 

These are the exact space-wise and time-wise computational complexity advantages that have the potential to significantly improve machine learning performance on quantum systems.

\subsection{Grover's Unstructured Search}
Grover's algorithm is a quantum algorithm that uses oracularization and amplitude amplification to achieve quadratic speedup in an unstructured search problem.\cite{Qiskit-Textbook} The algorithm begins by putting the qubit channels encoding the input space $|x\rangle$ in uniform superposition to construct a database in which to search: $H^{\otimes^N}|x\rangle$. If we measured the solution space right now, all database entries would have an equal likelihood of being flagged: $|s\rangle=\frac{1}{\sqrt N}\sum_{x=0}^{N-1}|x\rangle$. However, we want the solution state, $|\omega\rangle$, to have a higher probability of being flagged than the other database entries. To achieve this, we can define a unitary gate $U_\omega|x\rangle$ called the \textit{oracle} that returns $|x\rangle$ if $x\neq \omega$ and $-|x\rangle$ if $x=\omega$, effectively adding a negative phase to the solution states. On a practical level, the oracle encodes search criteria or a solution verification condition; however, Grover’s algorithm is agnostic to the content of the oracle and treats it as a queryable black-box that flags the winning state, allowing for the generality that lies at the crux of Grover’s unstructured search. In the plane spanned by the solution, or winning, state $|\omega\rangle$ and the uniform superposition state $|s\rangle$ with the winning state $|\omega\rangle$ removed and the complex vector rescaled defined as $|s'\rangle$, the superposition state $|s\rangle$ has a non-zero component in the winning state basis vector and a non-zero component in the superposition state basis vector since it contains the winning state. Applying the oracle gate to the superposition state, $U_\omega|s\rangle$, we effectively reflect the initial state $|s\rangle$ across the $|s'\rangle$-axis since the phase of the winning state component of the superposition is negated. Now, applying the gate $U_s=2|s\rangle\langle s|-1$ to $U_\omega |s\rangle$, we reflect the state $U_\omega |s\rangle$ about the uniform superposition vector $|s\rangle$ mapping the state to $U_sU_\omega|s\rangle$, a vector closer to the $|\omega\rangle$ basis state than $|s\rangle$, where $U_s$ is defined as the \textit{diffuser} gate. This double reflection operation $U_\omega U_s$ corresponds to a rotation in the $|\omega\rangle$-$|s'\rangle$ space toward $|\omega\rangle$ and is referred to as \textit{amplitude amplification}. The amplitude will be maximized after a number of rotations linear in the input size, since the larger the database size, the more non-solution entries as compared to solution entries, the smaller the angle between $|s\rangle$ and the $|s'\rangle$-axis, the more rotations are required to get the uniform superposition initial state to the winning state. Since the probability of measuring the solution state is the square of probability amplitude, the number of rotations maximizing the probability of measurement is $\mathcal O(\sqrt N)$. In this manner, Grover's algorithm achieves quadratic speedup in unstructured search.

\subsection{Related Works}
Given the domain of quantum information science is a relatively recent innovation as an application of quantum properties of matter, quantum machine learning (QML) is a largely unexplored subdomain that has only now begun to have practical implications with advancements in scaling qubit architectures. Furthermore, much of scholarly efforts in the field have been dedicated to implementing quantum analogues of natively classical learning paradigms rather than developing novel models that intrinsically take advantage of properties such as superposition and entanglement.\cite{dunjko_machine_2018} One means of such native quantum learning involves integrating Grover’s unstructured search into a QML model. Existing scholarship has explored two general means of integration: (1) using Grover's as an efficient quantum optimizer for QNN training / model parameter selection and (2) using a Grover's oracle as a shallow classifier. 

In 2013, Liu et al. proposed a single-hidden-layer feed-forward quantum neural network (SLFQNN) that amplifies the probability amplitude of the quantum neuron state with the highest performance function, implementing the first integration mode.\cite{liu_single-hidden-layer_2013} The SLFQNN is capable of learning non-linearities using radial Gaussian and Wavelet activation functions; however, it consists of a single neuron resulting in a shallow network incapable of effectively learning complex relationships and loses the benefit of quadratic speed-up post-training. 

Du et al., in 2021, proposed a Grover-search based quantum learning scheme for classification (GBLS) that effectively represents a classifier as a Grover's oracle, implementing the second integration mode.\cite{du_grover-search_2021} The model redirects the quadratic speed-up of Grover's search from efficient optimization to efficient propagation and performs well on simple classification tasks; however, the paradigm loses the benefits of a neural-network-based architecture such as modularity and complexity-encoding tunable hyperparameters. 

Both models individually demonstrate experimental potential as effective QML models;\cite{liu_single-hidden-layer_2013}\cite{du_grover-search_2021} however, perhaps there is means by which these integration modes can be syncretized.

\subsection{Motivation}

Oracularization is at the crux of Grover's search, where the oracle is effectively an encoding of the ruleset that satisfies the search criteria | a search problem solution verification condition. Conversely, input and output are taken in and used to generate a set of rules mapping from input to output encoded in the sets of weights and biases in a neural network. As such, if the output of a neural network is a set of rules and the input to Grover's is a set of rules, then theoretically the two functions can be composed such that a neural network can be inductively trained to learn a Grover's oracle from a set of database-hit examples. In this manner, we syncretize the two integration modes by extending the single quantum neuron\cite{liu_single-hidden-layer_2013} in the architecture to a deep quantum neural network (DQNN) with multiple hidden layers and multiple neurons per layer allowing for higher complexity in learned relationships, and embedding it \textit{within} Grover's unstructured search as a trainable oracle classifier\cite{du_grover-search_2021}. Thus, our proposed paradigm retains the modularity/complexity advantages of a neural network while gaining quadratic speed-up in network propagation taking advantage of the disjoint complementarity of the two integration modes to define a novel integration paradigm that coheres the benefit of both modes and decoheres the detriments. This is the motivation behind the machine learning paradigm explored by the inductive Grover's oracular quantum neural network (IGO-QNN) model.

\section{Model Architecture}
\subsection{Quantum Neurons \& Entangle Synapses}
The inductive Grover's oracular quantum neural network (IGO-QNN) model we propose can be abstracted as a directed quantum graph of \textit{quantum neural nodes} and \textit{entangle synaptic edges}. Quantum neurons are encoded as a $U_3(\theta,\phi,\lambda)$ gate, where the gate and its parameters parallel classical neuron bias, a quantum nonlinear activation function such as radial Gaussian basis or Morelet wavelet basis functions\cite{liu_single-hidden-layer_2013} encoded as a set of gates, and its outbound entangle-synapses, which connect it to other neurons. Entangle synapses are constituted of a trainable $U_3(\theta,\phi,\lambda)$ (distinct from the $U_3$ associated with a neural node) and a $CX$-gate which, when applied on qubits put in superposition via Hadamard gates, entangles two qubit channels by conditionally flipping the state of the target qubit depending on the state of the source qubit. The entangle synapse $U_3$-gate parameters serve as weights determining the extent to which two qubits are entangled, and can effectively allow for no entanglement of two qubits if nulled out ($L_1$ regularization can help fully null these weights). 

\begin{figure}[h]
    \centering
    \includegraphics[scale=0.45]{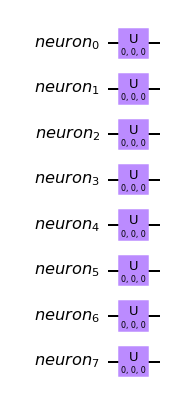}
    \includegraphics[scale=0.45]{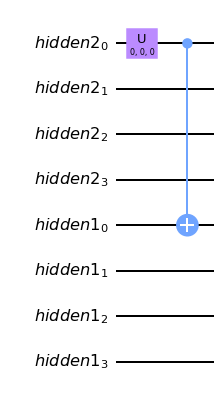}
    \caption{[Left] Quantum neuron qubit channels post parameter initialization. (8). [Right] Entangle synapse between quantum neurons hidden2$_0$ and hidden1$_0$.}
    \label{fig:synapse}
\end{figure}

\subsection{Network Layers}
The neural network consists of input, output, and hidden quantum layers, an oracle channel register, and a classical measure register. 

The input layer is a variational quantum register of $N$ qubit channels encoding unstructured database entries that is dynamically modified at each training iteration where a database-hit pair is passed as arguments to the network circuit. The output layer is a static quantum register of $N$ qubit channels encoding search hits and is the set of qubit channels are are measured and mapped to the classical output register of $N$ bits.

The $\ell$ hidden layers of the neural network are each constituted of $n_{\ell_i}$ neuron qubit channels. Each neuron in $\ell_i$ is densely connected to each neuron in $\ell_{i-1}$ for $1<i<\ell$ via an entangle synapse where neuron $n_i$ is the control neuron and $n_{i-1}$ is the target neuron. The gate abstracting this set of dense entangle synapse connections is referred to as the \textit{neural entangler}. The final hidden layer, $\ell_0$ is densely connected to each database entry in the output layer constituting the creation of the oracle in the output layer (oracle generator gate) that can then be applied to the input layer by another dense connection (oracularizer gate). Thus, the trainable neurons and synapses in the hidden layers conditionally applied to the output layer of the network effectively create a trainable Grover's unstructured oracle to be applied to $\ell_0$, where the number of hidden layers ($\ell$) and neurons per layer ($n_{\ell_i}$) encode the complexity of the search criteria.

\begin{figure}[h]
    \centering
    \includegraphics[width=\textwidth]{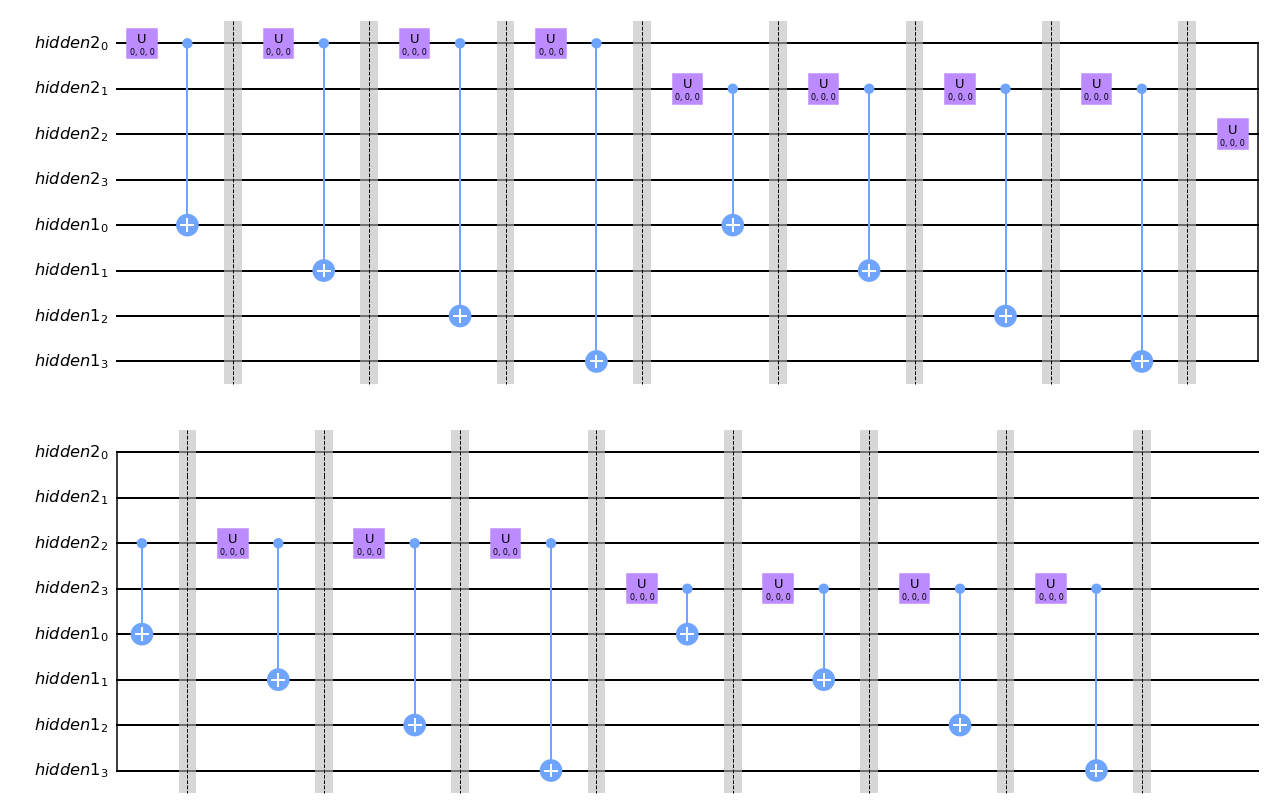}
    \caption{\textit{Neural entangler} operation between hidden layers $\ell_2$ and $\ell_1$. Entangle synapses densely connect each quantum neuron in $\ell_2$ (as control) with each in $\ell_1$ (as target).}
    \label{fig:neural_entangler}
\end{figure}

\subsection{Network Propgation Behavior}
The neural network is functionally encoded as a variational quantum circuit implementing a parameterized representation of Grover's search. The quantum neurons in each hidden layer and the oracle qubit are put in superposition via Hadamard transforms. Then the neural weights are initialized and applied to each neuron channel. Deeper hidden layer neurons first apply their respective activation functions and then are sequentially densely connected with each other via entangle synapses, and the final hidden layer is densely entangled with the output layer to create the oracle which is then applied to the input database. The output state of the neurally-encoded checker is then mapped to the oracle qubit (oracle flagging). Uncomputation using the inverse gate of the oracularizer is applied to the input and output layers, and the diffuser is applied for hit-state amplitude amplification. This sequence of oracularization, oracle flagging, uncomputation, and diffusion operations are iteratively applied $\sqrt N$ times (the optimal number of iterations for maximal amplitude amplification in Grover's algorithm). Finally, the oracle generation, neural entanglement, and weight initialization inverse gates are applied for uncomputation, and the output layer quantum register is measured to return the search hits for the oracle generated by the current model parameters. This network propagation behavior effectively allows for the neural network to search for hits in the input database using the trainable Grover's oracle. 

\subsection{Model Training}
The inductive Grover's oracular neural network is trained via a classical optimizer. At each training iteration, the variational (parameterized) circuit is provided input arguments ($X$-gates applied to represent 1 in the input data), the network propagates, and the returned hits are compared to the expected hits. This comparison yields a value of a classical loss function, and the set of all model parameters constituted of the $\theta$, $\phi$, and $\lambda$ for each $U_3$ gate representing each quantum neuron and entangle synapse weight, are tweaked via the optimizer (such as Gradient Descent or Adam) to minimize the loss. The classical optimizer interfaces with an OpenQASM abstraction of the IGO-QNN class where the trainable params attribute is the dictionary of classical parameter components (floating-point), mutable by the optimizing agent. The classical optimizer iteratively evaluates the model training performance until the loss function appears to have been minimized, then returns a static quantum circuit with the optimal parameters fixed into the quantum neurons and entangle synapses | this is the ideal Grover's oracle encoding the ruleset for the trained search task, which can be continually tuned in a live training paradigm to reduce sensitivity to perturbations in the database distribution of hits. 

\section{Applications}
The current paradigm for oracularization is the deductive representation of a decision problem (determining whether a proposed solution to a given problem is indeed a valid solution). As a consequence, an explicit solution to a given decision problem i.e. a solution-verification algorithm must be known in order to Grover's query a database for solutions to the associated search problem. This is the primary limitation of this deductive Grover's paradigm that narrows the range of applications to solely problems that contain closed form criteria for verification. However, of the set of all search problems, it is probability 1 that a given problem does not have a known deductive solution verifier. By means of the described quantum neural network architecture (IGO-QNN), we have described an inductive paradigm for the encoding of Grover's unstructured search oracles that allows for fast problem-space solution flagging without a closed form analytical framework for  verification. This widens the range of problem applications to include that vast majority without deductively-described oracles allowing for quadratic speed-up in unstructured search for problems with complex relationships between input and output spaces. The only prerequisite for the application of an IGO-QNN is that sufficient database-hit training examples constitute the training set for the classical optimizer to effectively minimize oracularization loss. 

This generalization of Grover's oracularization may prove particularly effective in deep reinforcement learning, computer vision, and, more generally, as a feature vector classifier at the top of an existing model. In the subdomain of reinforcement learning, a deep IGO-QNN can be given a database of potential future actions and the current state of the agent and flag the optimal decision. In the context of computer vision, an IGO-CNN can be given a database of images (likely semantically encoded into feature vectors) and be trained on matching image classification search hits to learn an oracle to flag images of a given classification. Any generalized set of feature vectors can be set as the input space of the network and the model can be trained to mark feature vectors that satisfy some condition for a general classification task. The network’s non-restrictive input/output space definition allows for it to be modularized and integrated as a layer in a larger machine learning model such as a hybrid quantum-classical context, allowing for the development of more robust models.

\section{Conclusion}
Thus, the inductive Grover oracular quantum neural network (IGO-QNN) theoretically significantly widens the range of problem applications of Grover's unstructured search algorithm to include the vast majority of problems lacking analytic descriptions of solution verifiers, allowing for quadratic speed-up in unstructured search for the set of search problems with relationships between input and output spaces that are tractably underivable deductively. Over a classical neural network, the model sports quadratic speed-up in network propagation using amplitude amplification, an exponentially more space-efficient neural representation using quantum superposition, and higher levels of encoded query complexity using quantum entanglement. Future research can implement and experimentally evaluate the performance of the machine learning model in different AI subdomains, particularly CV, NLP, RL, and neurosymbolic artificial intelligence, as well as with different classical, or potentially meta-quantum, optimizers (such as Grover’s unstructured search for optimal model parameters as demonstrated in the SLFQNN)\cite{liu_single-hidden-layer_2013}, loss functions, and activation functions. Furthermore, the oracular complexity is limited by the high number of qubit channels ($2N+\sum_{i=0}^\ell n_{\ell_i}+1$) required for the variational circuit representation of network | the simplest deep network requires 4 qubits; refining the space-wise efficiency of the model would be beneficial in this context. At a higher level of abstraction the paradigm of embedding a neural network within a variational Grover's search quantum circuit itself may be explored further to search for more nearly native quantum network parameter encodings. The model architecture as implemented in IBM Qiskit can be explored \href{https://colab.research.google.com/drive/1bDIG9gh6eqG8RDfFNQr00N3KxMR9XXct?usp=sharing}{here}, and is open-sourced.

\section{Acknowledgements}
I would like to thank Dr. Penman for his guidance in the research process and feedback regarding writing style as well as the incredible opportunity to explore both quantum computing and machine learning in the context of a writing seminar. I would like to thank George Zhou, Lu Esteban, and the Princeton Quantum Computing team as a whole for coming along for the ride exploring the vast world of quantum information science with me.

\section{References}
\printbibliography[heading=none]

@article{liu_single-hidden-layer_2013,
	title = {Single-hidden-layer feed-forward quantum neural network based on Grover learning},
	volume = {45},
	issn = {08936080},
	url = {https://linkinghub.elsevier.com/retrieve/pii/S0893608013000531},
	doi = {10.1016/j.neunet.2013.02.012},
	pages = {144--150},
	journaltitle = {Neural Networks},
	shortjournal = {Neural Networks},
	author = {Liu, Cheng-Yi and Chen, Chein and Chang, Ching-Ter and Shih, Lun-Min},
	urldate = {2021-04-16},
	date = {2013-09},
	langid = {english},
}

@article{cai_entanglement-based_2015,
	title = {Entanglement-Based Machine Learning on a Quantum Computer},
	volume = {114},
	issn = {0031-9007, 1079-7114},
	url = {http://arxiv.org/abs/1409.7770},
	doi = {10.1103/PhysRevLett.114.110504},
	pages = {110504},
	number = {11},
	journaltitle = {Physical Review Letters},
	shortjournal = {Phys. Rev. Lett.},
	author = {Cai, X.-D. and Wu, D. and Su, Z.-E. and Chen, M.-C. and Wang, X.-L. and Li, L. and Liu, N.-L. and Lu, Chao-Yang and Pan, Jian-Wei},
	urldate = {2021-04-16},
	date = {2015-03-19},
	langid = {english},
	eprinttype = {arxiv},
	eprint = {1409.7770},
}

@article{du_grover-search_2021,
	title = {A Grover-search based quantum learning scheme for classification},
	volume = {23},
	issn = {1367-2630},
	url = {https://iopscience.iop.org/article/10.1088/1367-2630/abdefa},
	doi = {10.1088/1367-2630/abdefa},
	pages = {023020},
	number = {2},
	journaltitle = {New Journal of Physics},
	shortjournal = {New J. Phys.},
	author = {Du, Yuxuan and Hsieh, Min-Hsiu and Liu, Tongliang and Tao, Dacheng},
	urldate = {2021-04-16},
	date = {2021-02-01},
	langid = {english},
}

@article{dunjko_machine_2018,
	title = {Machine learning \& artificial intelligence in the quantum domain: a review of recent progress},
	volume = {81},
	issn = {0034-4885, 1361-6633},
	url = {https://iopscience.iop.org/article/10.1088/1361-6633/aab406},
	doi = {10.1088/1361-6633/aab406},
	shorttitle = {Machine learning \& artificial intelligence in the quantum domain},
	pages = {074001},
	number = {7},
	journaltitle = {Reports on Progress in Physics},
	shortjournal = {Rep. Prog. Phys.},
	author = {Dunjko, Vedran and Briegel, Hans J},
	urldate = {2021-04-16},
	date = {2018-07-01},
	langid = {english},
}

@article{clark_basis_2014,
	title = {Basis for a neuronal version of Grover's quantum algorithm},
	volume = {7},
	issn = {1662-5099},
	url = {http://journal.frontiersin.org/article/10.3389/fnmol.2014.00029/abstract},
	doi = {10.3389/fnmol.2014.00029},
	journaltitle = {Frontiers in Molecular Neuroscience},
	shortjournal = {Front. Mol. Neurosci.},
	author = {Clark, Kevin B.},
	urldate = {2021-04-16},
	date = {2014-04-17},
	langid = {english},
}

@article{beer_training_2020,
	title = {Training deep quantum neural networks},
	volume = {11},
	issn = {2041-1723},
	url = {http://www.nature.com/articles/s41467-020-14454-2},
	doi = {10.1038/s41467-020-14454-2},
	pages = {808},
	number = {1},
	journaltitle = {Nature Communications},
	shortjournal = {Nat Commun},
	author = {Beer, Kerstin and Bondarenko, Dmytro and Farrelly, Terry and Osborne, Tobias J. and Salzmann, Robert and Scheiermann, Daniel and Wolf, Ramona},
	urldate = {2021-04-16},
	date = {2020-12},
	langid = {english},
}

@article{skolik_layerwise_2021,
	title = {Layerwise learning for quantum neural networks},
	volume = {3},
	issn = {2524-4906, 2524-4914},
	url = {http://link.springer.com/10.1007/s42484-020-00036-4},
	doi = {10.1007/s42484-020-00036-4},
	pages = {5},
	number = {1},
	journaltitle = {Quantum Machine Intelligence},
	shortjournal = {Quantum Mach. Intell.},
	author = {Skolik, Andrea and {McClean}, Jarrod R. and Mohseni, Masoud and van der Smagt, Patrick and Leib, Martin},
	urldate = {2021-04-16},
	date = {2021-06},
	langid = {english},
}

@electronic{Qiskit-Textbook,
	author = {Abraham Asfaw and Luciano Bello and Yael Ben-Haim and Mehdi Bozzo-Rey and Sergey Bravyi and Nicholas Bronn and Lauren Capelluto and Almudena Carrera Vazquez and Jack Ceroni and Richard Chen and Albert Frisch and Jay Gambetta and Shelly Garion and Leron Gil and Salvador De La Puente Gonzalez and Francis Harkins and Takashi Imamichi and Hwajung Kang and Amir h. Karamlou and Robert Loredo and David McKay and Antonio Mezzacapo and Zlatko Minev and Ramis Movassagh and Giacomo Nannicni and Paul Nation and Anna Phan and Marco Pistoia and Arthur Rattew and Joachim Schaefer and Javad Shabani and John Smolin and John Stenger and Kristan Temme and Madeleine Tod and Stephen Wood and James Wootton.},
	title = {Learn Quantum Computation Using Qiskit},
	year = {2020},
	url = {http://community.qiskit.org/textbook},
}

\begin{figure}[h]
    \centering
    \includegraphics[width=\textwidth]{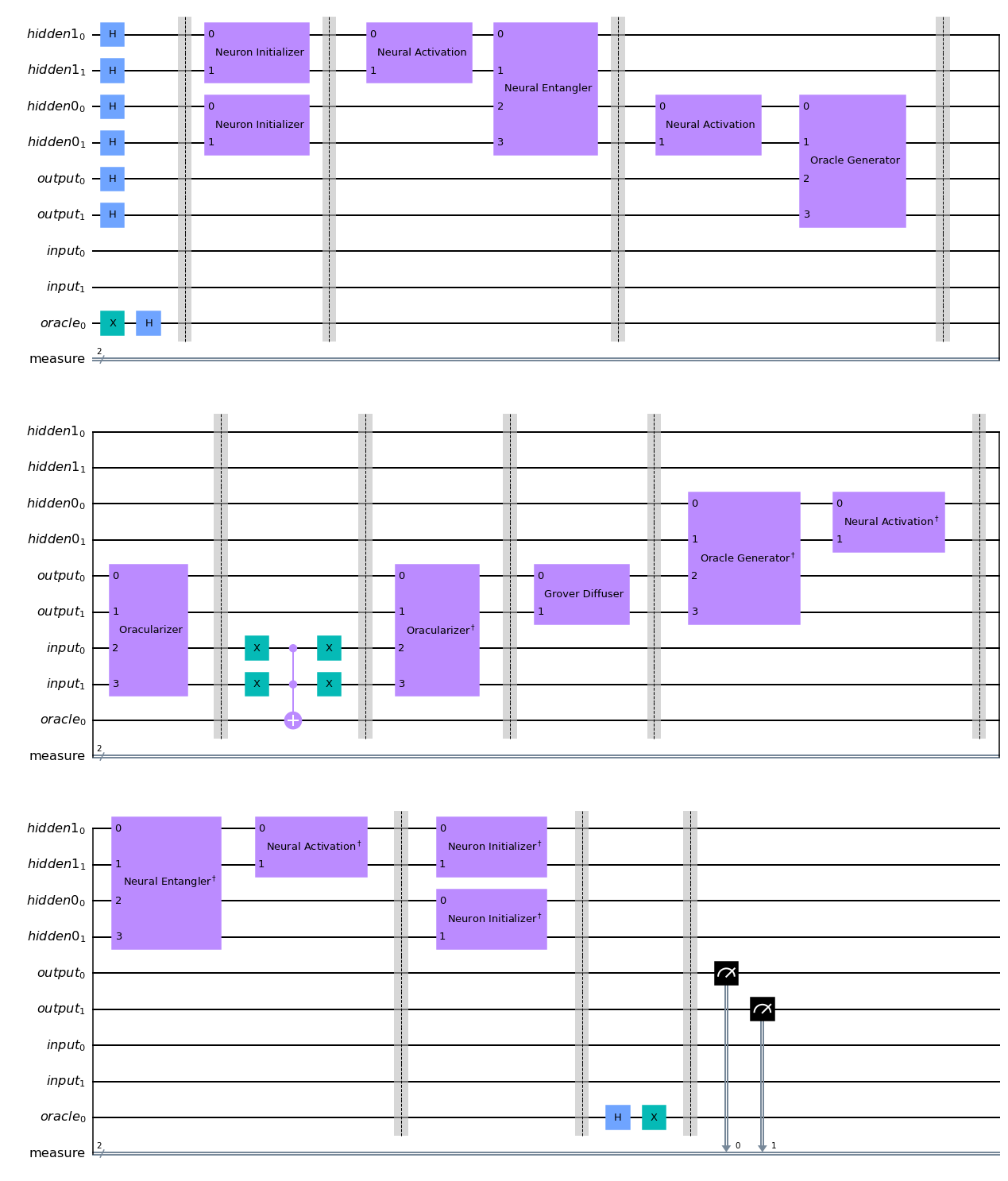}
    \caption{Full inductive Grover's oracular neural network (IGO-QNN) with structural parameters $N=2,\,\ell=2,\, n_{\ell_1}=n_{\ell_2}=4$ | database size of 2, two hidden layers of 4 quantum neurons each.}
    \label{fig:full_circuit}
\end{figure}

\begin{figure}[h]
    \centering
    \includegraphics[scale=0.2]{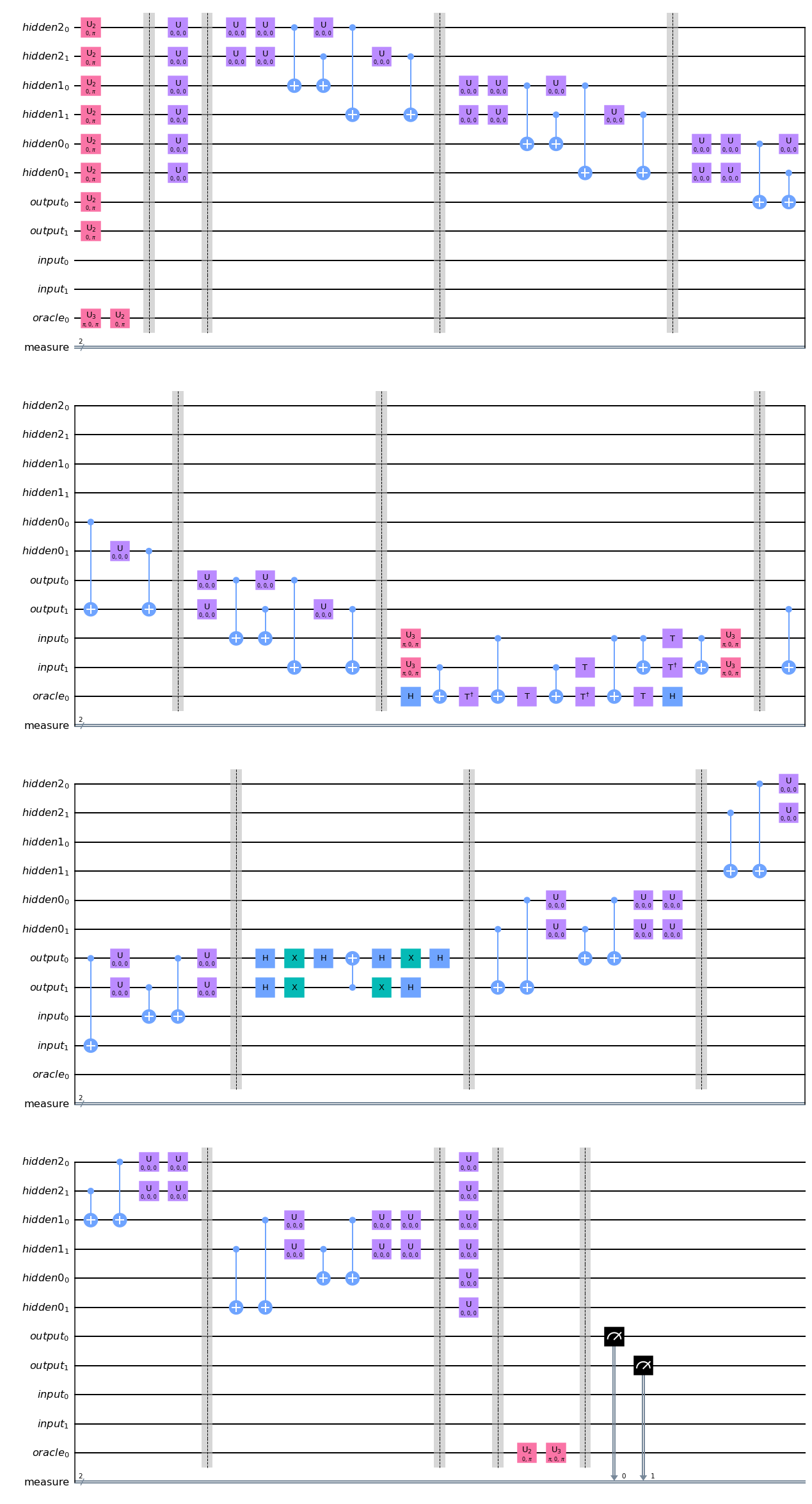}
    \caption{Decomposed IGO-QNN with structural parameters $N=2,\,\ell=2,\, n_{\ell_1}=n_{\ell_2}=4$ | database size of 2, two hidden layers of 4 quantum neurons each.}
    \label{fig:decomposed_circuit}
\end{figure}

\end{document}